\documentclass[a4paper]{article}
\usepackage{amsmath,amssymb}
\usepackage{graphicx}

\newtheorem{theorem}{Theorem}

\newtheorem{ex}{Example}
\newtheorem{rem}{Remark}

\def\rmi{\mathrm{i}}
\def\rme{\mathrm{e}}
\def\rmd{\mathrm{d}}

\def\im{\mathrm{Im\,}}

\def\openone{\leavevmode\hbox{\small1\kern-3.3pt\normalsize1}}
\def\diag{\mbox{diag\,}}

\def\bbbc{\mathbb{C}}
\def\bbbr{\mathbb{R}}
\def\bbbz{\mathbb{Z}}

\begin{document}

\title{Dressing method and quadratic bundles related to symmetric spaces.
Vanishing boundary conditions}%
\author{T. I. Valchev\footnote{On leave of the Institute for Nuclear Research and Nuclear
Energy, Bulgarian Academy of Sciences, 72 Tsarigradsko chaussee Blvd., 1784 Sofia, Bulgaria,}\\
\small School of Mathematical Sciences,\\
\small Dublin Institute of Technology,\\
\small Kevin Street, Dublin 8, Ireland}
\date{29-th September, 2014}

\maketitle

\begin{abstract}
We consider quadratic bundles related to Hermitian symmetric spaces of the type $\mathrm{SU}(m+n)/\mathrm{S}(\mathrm{U}(m)\times\mathrm{U}(n))$.
The simplest representative of the corresponding integrable hierarchy is given by a multi-component Kaup-Newell derivative nonlinear
Schr\"odinger equation which serves as a motivational example for our general considerations. We extensively discuss how one can apply
Zakharov-Shabat's dressing procedure to derive reflectionless potentials obeying zero boundary conditions. Those could be used for one to
construct fast decaying solutions to any nonlinear equation belonging to the same hierarchy. One can distinguish between generic soliton type
solutions and rational solutions. 
\end{abstract}

\section{Introduction}

A classical example of a completely integrable nonlinear evolution equation (NLEE) is derivative nonlinear Schr\"odinger equation (DNLS $"\pm"$) 
\begin{equation}
\rmi q_{t} + q_{xx} \pm \rmi \left(q^2 q^*\right)_x =0, 
\label{dnls}
\end{equation}
where the subscripts denote partial differentiation in variables $t$ and $x$ and $*$ stands  for complex conjugation. DNLS finds applications
in plasma physics\cite{mio,mjol,rud1,rud2}. It was Kaup and Newell \cite{kaupnewel} who has shown DNLS has a Lax pair of the form:
\begin{eqnarray}
L(\lambda) &=& \rmi \partial_x + \lambda Q(x,t) - \lambda^2\sigma_3\label{lax1_scalar}\\
A(\lambda) &=&\rmi\partial_t + \sum^3_{k=1} A_k(x,t) \lambda^k - 2\lambda^4\sigma_3
\end{eqnarray}
where $\lambda\in\bbbc$ is spectral parameter and
\begin{eqnarray*}
Q &=& \left(\begin{array}{cc}
0 & q \\ \pm q^* & 0
\end{array}\right),\qquad \sigma_3 = \left(\begin{array}{cc}
1 & 0 \\ 0 & -1
\end{array}\right),\qquad A_3 = 2Q\\
A_2 &=& \pm |q|^2\sigma_3,\qquad A_1 = \frac{\rmi}{2}[\sigma_3,Q_x]
\pm |q|^2 Q.	
\end{eqnarray*}
DNLS is tightly related to three other integrable NLEEs. These are Chen-Lee-Liu's equation \cite{cll} 
\begin{equation}
\rmi q_t + q_{xx} +\rmi qq^*q_x = 0,
\label{dnls2}
\end{equation}
Gerdjikov-Ivanov's equation \cite{gerdjivan2} 
\begin{equation}
\rmi q_{t} + q_{xx} + \rmi q^2 q^*_x + \frac{1}{2}|q|^4q =0
\label{dnls3}
\end{equation}
and 2-dimensional Thirring model \cite{mikthir}
\begin{eqnarray}
\left(\rmi\partial_{\sigma}\gamma^{\sigma} - m\right)\psi 
\pm g \gamma^{\nu}\psi \psi^{\dag}\gamma^{0}\gamma_{\nu}\psi = 0,\qquad
\sigma,\nu = 0,1\\	
\gamma^0 = \left(\begin{array}{cc}
0 & 1 \\ 1 & 0
\end{array}\right),\qquad \gamma^1 = \left(\begin{array}{cc}
0 & -1 \\ 1 & 0
\end{array}\right)
\end{eqnarray}
where $\psi:\bbbr^2\to\bbbc^2$ is a smooth spinor field, $g$ is a bonding constant and $\dag$ means Hermitian conjugation (Einstein's rule
of summation over repeated indices holds above).

Due to their similarity with DNLS, (\ref{dnls2}) and (\ref{dnls3})
are sometimes termed DNLS II and DNLS III respectively. All the three DNLS versions along with the 2-dimensional
Thirring model correspond to certain Mikhailov type of reductions of the generic quadratic bundle operator \cite{gerdjivan2, mikthir}:
\begin{equation}
L(\lambda) = \rmi \partial_x + U_0(x,t)  + \lambda U_1(x,t) - \lambda^2\sigma_3\\
\label{cqb}
\end{equation}
where $U_1(x,t)$ is an off-diagonal $2\times 2$ matrix and $U_{0}(x,t)$ being a traceless
$2\times 2$ matrix otherwise arbitrary. 

One very fruitful trend in theory of integrable systems is search and study of multi-component counterparts of scalar completely
integrable equations\cite{AthFor, ForKu*83, ours, nls, book}. Fordy and Kulish \cite{ForKu*83} proposed a natural way to relate to
each Hermitian symmetric space a multi-component nonlinear Schr\"odinger equation. Later Fordy\cite{fordy} managed to find similar
connection between Hermitian symmetric spaces and multi-component versions of DNLS (\ref{dnls}).  For example, the equation 
\begin{equation}
\rmi \mathbf{q}_{t} + \mathbf{q}_{xx} + \frac{2m\rmi}{m + n}\left(\mathbf{q}\mathbf{q}^{\dag}\mathbf{q}
\right)_x = 0\label{mdnls}	
\end{equation}
where $\mathbf{q}$ is a smooth $n\times m$ matrix-valued function is related to symmetric space
$\mathrm{SU}(m+n)/\mathrm{S}(\mathrm{U}(m)\times\mathrm{U}(n))$. Similarly, Tsuchida and Wadati \cite{wadcll}
proved the complete integrability of a matrix generalization of Chen-Lee-Liu equation. 

We aim here at demonstrating how one can adapt Zakharov-Shabat's dressing method \cite{brown-bible, zahshab}
to incomplete quadratic bundles of the type
\begin{eqnarray}
L(\lambda) & = & \rmi\partial_x + \lambda Q(x,t) - \lambda^2 J \label{qb_a3_0}\\
Q(x,t)  &=&  \left(\begin{array}{cc}
0 & \mathbf{q}^T(x,t)\\ 
\mathcal{E}_{n}\mathbf{q}^*(x,t)\mathcal{E}_{m} & 0
\end{array}\right),\qquad
J=\left(\begin{array}{cc}
\frac{n}{m}\openone_m & 0 \\
0 & -\openone_n
\end{array}\right)\nonumber
\end{eqnarray}
where $\mathcal{E}_{m}$ is a diagonal matrix of dimension $m$ respectively with diagonal entries equal to $\pm 1$
while $\openone_{m}$ is the unit matrix of dimension $m$. This operator is tightly related to symmetric spaces of
the form $\mathrm{SU}(m+n)/\mathrm{S}(\mathrm{U}(m)\times\mathrm{U}(n))$ and its quadratic flow produces equation
\begin{equation}
\rmi \mathbf{q}_{t} + \mathbf{q}_{xx} + \frac{2m\rmi}{m+n}
\left(\mathbf{q}\mathcal{E}_{m}\mathbf{q}^{\dag}\mathcal{E}_{n}\mathbf{q}\right)_x = 0
\label{mdnls_pseudo_0}	
\end{equation}
which generalizes Fordy's equation (\ref{mdnls}). We shall restrict ourselves with potentials obeying zero boundary
condition
\begin{equation}
\lim_{x\to\pm\infty} Q(x,t) = 0.
\end{equation}
Those give rise to fast decaying solutions to NLEEs belonging to the integrable hierarchy generated
by (\ref{qb_a3_0}).

The paper is structured as follows. Second section is preliminary in nature. It contains a
brief summary of some basic properties of the quadratic bundles related to symmetric spaces of the type
$\mathbf{A.III}$ and the auxiliary linear problem associated with it. The following two sections contain
our main results. Section \ref{dres} contains general considerations of Zakharov-Shabat's dressing method.
We show how one can adapt the dressing method for quadratic bundles of the afore-mentioned type.
This allows one to obtain reflectionless potentials and thus generate special types of solutions on a
trivial (zero) background in a purely algebraic manner. We shall see there are two different types of
solutions: generic soliton type solutions and rational solutions. The soliton type solutions are
associated with dressing factors whose poles are generic while the rational solutions correspond to factors
whose poles lie on the continuous spectrum of (\ref{qb_a3_0}). In section \ref{sol} we explicitly construct
reflectionless potentials and particular solutions of either of the afore-mentioned types. The latter can be
reduced to well-known solutions to the scalar DNLS "$\pm$" as a very special case. Last section contains summary
of our results and some additional remarks.

\section{Quadratic bundles and symmetric spaces of the type $\mathbf{A.III}$}\label{qb}

In this section we shall introduce some basic notions of direct scattering problem for quadratic bundles
related to symmetric spaces of the type $\mathrm{SU}(m+n)/\mathrm{S}(\mathrm{U}(m)\times \mathrm{U}(n))$,
and sketch some of their properties. In doing this we are going to use some results and conventions from
\cite{gerdjivan, book, varna13}.

Let us consider the following Lax pair:
\begin{eqnarray}
L(\lambda) &=& \rmi\partial_x + \lambda Q(x,t) - \lambda^2 J,
\label{lax_1}\\
A(\lambda) &=& \rmi\partial_t + \sum^{2N}_{k=1}\lambda^k A_k(x,t)\label{lax_2}
\end{eqnarray}
where $\lambda\in\bbbc$ is spectral parameter while $Q(x,t)$, $J$ and $A_k(x,t)$ are $(m+n)\times(m+n)$
matrices. Further we shall denote by $\mathrm{M}_{m,n}[\bbbc]$ linear space of all $m\times n$ matrices
with complex entries. All coefficients are assumed to be traceless matrices to satisfy the
following Mikhailov type reduction conditions \cite{miktetr, mik} 
\begin{eqnarray}
\mathcal{E}Q^{\dag}\mathcal{E} = Q,\qquad	\mathcal{E}J^{\dag}\mathcal{E} = J,\label{pseudoherm}\\
\mathcal{E}A^{\dag}_{k}\mathcal{E} = A_{k},\qquad \mathcal{E} = \diag(\epsilon_{1},\ldots,\epsilon_{m+n}),
\end{eqnarray}
for $\epsilon^2_1 =\ldots =\epsilon^2_{m+n} = 1$.
To relate the Lax pair (\ref{lax_1}) and (\ref{lax_2}) with a symmetric
space of the type of $\mathbf{A.III}$, $L$ and $A$ are required to obey the following constraints:
\begin{equation}
\mathbf{C}L(-\lambda)\mathbf{C} = L(\lambda),\qquad
\mathbf{C}A(-\lambda)\mathbf{C}=A(\lambda).
\label{red_LA}
\end{equation}
The constant matrix $\mathbf{C} = \diag(\openone_m,-\openone_n)$ is connected to Cartan's involution \cite{hel} defining
Hermitian symmetric space $\mathrm{SU}(m+n)/\mathrm{S}(\mathrm{U}(m)\times\mathrm{U}(n))$. It induces a $\bbbz_2$ grading
in Lie algebra $\mathfrak{sl}(m+n,\bbbc)$, namely we have:
\begin{equation}
\mathfrak{sl}(m+n) = \mathfrak{sl}^{0}(m+n) + \mathfrak{sl}^{1}(m+n)
\label{sl_split}
\end{equation}
where
\[\mathfrak{sl}^{\sigma}(m+n) = \{X\in\mathfrak{sl}(m+n)|\; \mathbf{C}X\mathbf{C}=(-1)^{\sigma}X\}, \qquad \sigma = 0,1\]
are eigen subspaces of the adjoint action of $\mathbf{C}$.
Due to (\ref{red_LA}) $Q$ acquires block off-diagonal structure:
\begin{eqnarray}
Q  &=&  \left(\begin{array}{cc}
0 & \mathbf{q}^T\\
\mathcal{E}_{n}\mathbf{q}^*\mathcal{E}_{m} & 0
\end{array}\right),\label{q_a3}\qquad
\mathcal{E}_{m} = \diag(\epsilon_{1},\ldots,\epsilon_{m})\\
\mathcal{E}_{n} & = & \diag(\epsilon_{m+1},\ldots,\epsilon_{m+n})\nonumber
\end{eqnarray}
for $\mathbf{q}: \bbbr^2\to\mathrm{M}_{n,m}[\bbbc]$ while $J$ is a
block-diagonal matrix. Similarly, $A_{2l-1}$ and $A_{2l}$ for
$l=1,\ldots,N$ acquire the block form
\begin{eqnarray}
A_{2l-1} & = & \left(\begin{array}{cc}
0 & \mathbf{a}^T_{2l-1}\\
\mathcal{E}_{n}\mathbf{a}^*_{2l-1}\mathcal{E}_{m} & 0
\end{array}\right),\\
A_{2l} & = & \left(\begin{array}{cc}
\mathbf{a}_{2l}\\
0 & \mathbf{b}_{2l}
\end{array}\right)
\label{a_2l_a3}
\end{eqnarray}
where $\mathbf{a}_{2l-1}: \bbbr^2\to\mathrm{M}_{n,m}[\bbbc]$, $\mathbf{a}_{2l}: \bbbr^2\to \mathrm{M}_{m,m}[\bbbc]$
and $\mathbf{b}_{2l}: \bbbr^2\to \mathrm{M}_{n,n}[\bbbc]$. In addition, $\mathbf{a}_{2l}$ and $\mathbf{b}_{2l}$
obey the symmetries:
\begin{equation}
\mathcal{E}_{m}\mathbf{a}^{\dag}_{2l}\mathcal{E}_{m} = \mathbf{a}_{2l},
\qquad\mathcal{E}_{n}\mathbf{b}^{\dag}_{2l}\mathcal{E}_{n} = \mathbf{b}_{2l}.
\end{equation}
For convenience we shall pick up the constant matrix $J$ in the form
\begin{equation}
J=\left(\begin{array}{cc}
\frac{n}{m}\openone_m & 0 \\
0 & -\openone_n
\end{array}\right).
\label{car_el}
\end{equation}
Its centralizer coincides with $\mathfrak{sl}^{0}(m+n)$.

The quadratic flow, that is $N=2$, for the Lax pair (\ref{lax_1}) and (\ref{lax_2}) produces the following
multi-component DNLS
\begin{equation}
\rmi \mathbf{q}_{t} + \mathbf{q}_{xx} + \frac{2m\rmi}{m+n}
\left(\mathbf{q}\mathcal{E}_{m}\mathbf{q}^{\dag}\mathcal{E}_{n}\mathbf{q}\right)_x = 0.
\label{mdnls_pseudo}	
\end{equation}
Clearly, equation (\ref{mdnls_pseudo}) represents a natural generalization of Fordy's equation (\ref{mdnls})
for it includes DNLS "-" as a special scalar case.

Let us now consider the auxiliary linear problem
\begin{equation}
L(\lambda)\Psi(x,t,\lambda) = \rmi\partial_x\Psi
+ \lambda (Q - \lambda J)\Psi = 0
\label{aux_lin}\end{equation}
where $\Psi$ is a fundamental set of solutions (fundamental solution for short) hence $\det\Psi(x,t,\lambda)
\neq 0$ for any $x$, $t$ and $\lambda$ in its domain. We shall assume from now on that $Q$ is infinitely
smooth and obeys the boundary condition
\[\lim_{x\to\pm\infty}Q(x,t)= \mathbf{0}.\]
Following \cite{book} one defines Jost fundamental solutions $\Psi_{+}$ and $\Psi_{-}$ as follows:
\begin{equation}
\lim_{x\to\pm\infty}\Psi_{\pm}(x,t,\lambda)\rme^{\rmi\lambda^2 Jx}=\openone.
\label{josts}
\end{equation}
The transition matrix 
\[T(t,\lambda) = \hat{\Psi}_{+}(x,t,\lambda)\Psi_{-}(x,t,\lambda),\qquad 
\hat{\Psi} \equiv \Psi^{-1}\]
between the Jost solutions defines scattering matrix.
Since $[L, A] = 0$ any fundamental solution also fulfills linear system
\begin{equation}
\rmi\partial_t\Psi + \sum^{2N}_{k=1}\lambda^kA_k\Psi = \Psi f
\label{aux_lin2}
\end{equation}
where polynomial
\[f(\lambda) = \lim_{x\to\pm\infty}\sum^{2N}_{k=1}\lambda^kA_k(x,t)\]
is dispersion law of NLEE and it carries all its essential characteristics.
It can be proven \cite{book} that the scattering matrix evolves with time according to:
\[T(t,\lambda)=\rme^{\rmi f(\lambda)t}T(0,\lambda)\rme^{-\rmi f(\lambda)t} .\]
For convenience we shall omit variables $x$ and $t$ where this does not lead to confusion.

The Jost solutions are defined for real and imaginary values of $\lambda$ only. Starting from the Jost solutions,
however, one is able to construct another pair of solutions of (\ref{aux_lin}) to have analytic properties in upper
half plane and the lower half plane in $\lambda^2$-plane. More specifically, the following theorem holds true
\cite{gerdjivan, varna13}: 

\begin{theorem} [Gerdjikov\&Ivanov, 1983]

There exists a pair of solutions $X^{+}$ and $X^{-}$ analytic in domains $\Omega_{+} = \{\lambda\in \bbbc |
\,\im \lambda^2\geq 0\}$ (i.e. the first and the third quadrant in the $\lambda$-plane) and
$\Omega_{-} = \{\lambda\in \bbbc |\, \im \lambda^2\leq 0\}$ (the second and the forth quadrants resp.).
$X^{+}$ and $X^{-}$ can be constructed from the Jost solutions as follows:
\begin{equation}
X^{\pm}(\lambda) = \left\{\begin{array}{l}\Psi_{-}(\lambda)S^{\pm}(\lambda)\\
\Psi_{+}(\lambda)T^{\mp}(\lambda)D^{\pm}(\lambda)\end{array}\right.  .
\label{chi_constr}
\end{equation}
where matrices $S^{\pm}(\lambda)$, $T^{\pm}(\lambda)$ and $D^{\pm}(\lambda)$ are given by
\begin{eqnarray*}
S^{+}(\lambda) &=&  \left(\begin{array}{cc}
\openone_m & \mathbf{s}_+^T(\lambda) \\ \mathbf{0} & \openone_{n}
\end{array}\right),\qquad T^{+}(\lambda) =
\left(\begin{array}{cc}
\openone_m & \mathbf{t}^T_{+}(\lambda) \\ \mathbf{0} & \openone_{n}
\end{array}\right)\\
S^{-}(\lambda) &=&  \left(\begin{array}{cc}
\openone_m & \mathbf{0} \\ \mathbf{s}_{-}(\lambda) & \openone_{n}
\end{array}\right),\qquad T^{-}(\lambda) =
\left(\begin{array}{cc}
\openone_m & \mathbf{0}  \\ \mathbf{t}_{-}(\lambda)& \openone_{n}
\end{array}\right)\\
D^{\pm}(\lambda) &=&  \left(\begin{array}{cc}
d^{\pm}_m(\lambda) & \mathbf{0} \\ \mathbf{0} & d^{\pm}_n(\lambda)
\end{array}\right).
\end{eqnarray*}
The latter are involved in block LDU decomposition 
\begin{equation}
T(\lambda)=T^{\mp}(\lambda)D^{\pm}(\lambda)\hat{S}^{\pm}(\lambda))
\label{ldu}
\end{equation}
of the scattering matrix. The decomposition (\ref{ldu}) respects the splitting (\ref{sl_split}),
i.e. $d^{\pm}_n(\lambda)$ are $n\times n$ matrices, while $\mathbf{s}_{\pm}(\lambda)$ and
$\mathbf{t}_{\pm}(\lambda)$ are $n\times m$ matrices.\end{theorem}

The reductions imposed on the Lax operators yield to certain constraints on the
values of the Jost solutions, scattering matrix and fundamental analytic solutions
\cite{miktetr,mik}, namely we have:
\begin{equation}
\begin{split}
\hat{\Psi}^\dag_{\pm}(\lambda^*) & = \Psi_{\pm}(\lambda),\qquad
\mathbf{C}  \Psi_{\pm}(-\lambda)\mathbf{C}  = \Psi_{\pm}(\lambda)\\
\hat{T}^\dag (\lambda^*) & = T(\lambda),\qquad
\mathbf{C}  T(-\lambda)\mathbf{C} = T(\lambda)\\
\left[X^{+}(\lambda^*)\right]^{\dag} & = \hat{X}^{-}(\lambda), \qquad
\mathbf{C}  X^{\pm}(-\lambda)\mathbf{C} = X^{\pm}(\lambda).
\end{split}
\label{jost_fas_red}\end{equation}

It follows from (\ref{chi_constr}) the fundamental analytic solutions $X^{+}$ and
$X^{-}$ are interrelated through:
\begin{equation}
X^{+}(\lambda) = X^{-}(\lambda) G(\lambda),\qquad \lambda^2\in\bbbr
\label{rhp}
\end{equation}
for some sewing function $G(\lambda)=\hat{S}^{-}(\lambda)S^{+}(\lambda)$. This means that they can be
viewed as solutions to a local Riemann-Hilbert problem \cite{gerdjivan, book, brown-bible} with boundary given
by the real and imaginary lines in the $\lambda$-plane. More precisely, solutions to a local Riemann-Hilbert
problem are functions $\Upsilon^{\pm} = X^{\pm}\exp(\rmi\lambda^2Jx)$ satisfying the linear system
\begin{equation}
\rmi\partial_x\Upsilon^{\pm} + \lambda Q\Upsilon^{\pm} -\lambda^2 [J,\Upsilon^{\pm}] =0.
\label{ad_lin_sys}
\end{equation}
More detailed analysis of (\ref{ad_lin_sys}) shows that $\Upsilon^{+}$ and $\Upsilon^{-}$
are normalized as follows:
\begin{equation}
\lim_{\lambda\to\, 0} \Upsilon^{\pm}(x,\lambda) = \openone,	
\end{equation}
that is the Riemann-Hilbert problem is normalized at $\lambda = 0$.  

The fundamental analytic solutions can be used to describe the spectrum of the scattering
operator \cite{gerji, book}. More specifically, one can prove \cite{gerdjivan, varna13} the following theorem
holds true.
\begin{theorem}
The spectrum of $L(\lambda)$ comprises a continuous and discrete part. 
The continuous part of spectrum is determined by the condition:
\begin{equation}
\im \lambda^2 J = 0,
\label{cont_spec}\end{equation} 
i.e. it coincides with the real and the imaginary lines in the $\lambda$-plane.
The discrete spectrum belongs to (discrete) orbits of the reduction group $\bbbz_2\times\bbbz_2$,
i.e. all discrete eigenvalues go together in quadruples of points $\{\pm\mu, \pm\mu^*\}$
located symmetrically to the real and imaginary lines. $\Box$
\end{theorem}

\section{Dressing method}\label{dres}

In this section we are going to demonstrate how one can construct special solutions to any member
of the integrable hierarchy generated by Lax operator (\ref{lax_1}). For that purpose we shall employ Zakharov-Shabat's dressing
method adapted for quadratic bundles of the type discussed in the previous section. Let us start with a few
general remarks. 

Zakharov-Shabat's dressing method is an indirect way of integration of S-integrable equations
\cite{book, brown-bible, zahshab}, i.e. it generates new solutions to a given NLEE starting from
a known one. Its application is substantially determined by the existence and form of the Lax
representation associated with the NLEE. Suppose $\Psi_0$ is a fundamental solution to the system
\begin{eqnarray}
L_0(\lambda) \Psi_0 & = & \rmi\partial_x\Psi_0 + \lambda\left(Q^{(0)} - \lambda J\right)\Psi_0 = 0
\label{bare_x}\\
A_0(\lambda)\Psi_0 & = & \rmi\partial_t\Psi_0 +
\sum^{2N}_{k=1}\lambda^kA^{(0)}_k\Psi_0 = \Psi_0 f(\lambda)\label{bare_t}
\end{eqnarray}
to be referred to further on as bare system. The matrix coefficients above, having the same structure
as in (\ref{q_a3}), (\ref{a_2l_a3}), are assumed to be known. Let us now apply a gauge (dressing) transform
$\Psi_0 \to \Psi_1 = g\Psi_0$ such that the auxiliary linear systems remain covariant, i.e. $\Psi_1$
must satisfy 
\begin{eqnarray}
L_1(\lambda)\Psi_1 & = & \rmi\partial_x\Psi_1 + \lambda\left(Q^{(1)} - \lambda J\right)\Psi_1 = 0 \label{dres_x}\\
A_1(\lambda)\Psi_1 & = & \rmi\partial_t\Psi_1 + \sum^{2N}_{k=1}\lambda^kA^{(1)}_k\Psi_1 = \Psi_1 f(\lambda)
\label{dres_t}
\end{eqnarray}
for some new, unknown coefficients $Q^{(1)}$, $A^{(1)}_k$ of the same form as those in (\ref{bare_x}) and (\ref{bare_t}).
After comparing the bare system (\ref{bare_x}), (\ref{bare_t}) with the dressed one (\ref{dres_x}), (\ref{dres_t}) we see
that the dressing factor $g$ is a solution to the following pair of PDEs:
\begin{eqnarray}
\rmi\partial_x g + \lambda Q^{(1)}\,g - \lambda gQ^{(0)} - \lambda^2[J,g] & = & 0
\label{g_pde1}\\
\rmi\partial_t g +  \sum^{2N}_{k=1}\lambda^kA^{(1)}_k g - g \sum^{2N}_{k=1}\lambda^kA^{(0)}_k & = & 0.
\label{g_pde2}
\end{eqnarray}
In order to determine possible forms of $g$ regarding the spectral parameter $\lambda$ we analyze equations (\ref{g_pde1})
and (\ref{g_pde2}). Suppose $g$ does not depend on $\lambda$ then it is straightforward from (\ref{g_pde1}) and (\ref{g_pde2})
that it is simply a constant matrix, that is trivial. Thus to obtain a non-trivial result we shall require that $g$ does depend
on the spectral parameter. 

Further, we recall that bare fundamental solutions $\Upsilon^{\pm}_0$ and their dressed counterparts $\Upsilon^{\pm}_1$ satisfy
a Riemann-Hilbert problem with a normalization at $\lambda = 0$. This implies that $g$ is also normalized at $\lambda = 0$,
i.e. we have:
\begin{equation}
g(\lambda = 0) = \openone.	
\label{g_atzero}\end{equation}
On the other hand, due to constraints (\ref{jost_fas_red}) the dressing factor obeys the symmetry conditions:
\begin{eqnarray}
\mathcal{E}g^{\dag}(\lambda^*)\mathcal{E} & = & \hat{g}(\lambda)\label{g_red1}\\
\mathbf{C} g(-\lambda)\mathbf{C}  &=& g(\lambda).\label{g_red2}
\end{eqnarray}
A simple choice for the dressing factor to respect (\ref{g_atzero})--(\ref{g_red2}) is the following one:
\begin{equation}
g(x,t,\lambda) = \openone + \sum^{r}_{j=1} \frac{\lambda}{\mu_j}\left(\frac{B_j(x,t)}{\lambda-\mu_j}
+ \frac{\mathbf{C}B_j(x,t)\mathbf{C}}{\lambda + \mu_j} \right).
\label{g_multi}
\end{equation}

Equation (\ref{g_pde1}) allows one to find a simple interrelation between the bare potential $Q_0$ and the dressed one
$Q_1$. Indeed, after dividing both hand-sides of (\ref{g_pde1}) by $\lambda$, setting $|\lambda| \to \infty$ and taking
into account the form of $g$ and relation (\ref{g_red1}) we obtain:
\begin{equation}
Q^{(1)} = Q^{(0)} + \sum^{r}_{i = 1}[J, B_i - \mathbf{C}B_i\mathbf{C}]\mathcal{E} g^{\dag}_{\infty}\mathcal{E}
\label{q_1_q_0}
\end{equation}
where
\begin{equation}
g_{\infty}(x,t) = \lim_{|\lambda|\to\infty} g(x,t,\lambda).
\end{equation}
Thus to obtain a new solution we need to know the residues of the dressing factor. As it turns out the latter can
be expressed in terms of the bare fundamental solution $\Psi_0$ (and its first $\lambda$-derivative). This fact
constitutes the power of the dressing method.

In order to find the residues of $g$ one analyzes the identity $g\hat{g}=\openone$ and PDEs (\ref{g_pde1}) and
(\ref{g_pde2}). In what follows we shall distinguish between two different cases:
\begin{enumerate}
\item generic case, that is the poles of $g$ lie outside of the continuous spectrum of $L$;
\item degenerate case, i.e. the poles of $g$ lie on the continuous spectrum ($\mu^2_j\in\bbbr$).
\end{enumerate}

\subsection{Generic case} \label{dres1}

Let us consider first the case when the poles of $g$ and its inverse represent distinct points in $\lambda$-plane
symmetrically located with respect to the real and imaginary lines. Then after evaluating the residue of $g\hat{g}$
at $\lambda = \mu_i$, $i=1,\ldots,r$ we obtain the following algebraic relations:
\begin{equation}
B_i\left[\openone + \sum^{r}_{j=1}\frac{\mu_i}{\mu^*_j}\left(\frac{\mathcal{E} B^{\dag}_j\mathcal{E}}{\mu_i - \mu^*_j}
+ \frac{\mathcal{E}\mathbf{C} B^{\dag}_j\mathbf{C}\mathcal{E}}{\mu_i + \mu^*_j}\right)\right] = 0.
\label{b_rel}
\end{equation}
To ensure the result is nontrivial one needs to assume the residues are degenerate matrices \cite{varna13}, i.e. they
obey decomposition
\begin{equation}
B_i = X_iF^T_i
\label{Bi_decomp}
\end{equation}
for some rectangular matrices $X_i(x,t)$ and $F_i(x,t)$. After substituting (\ref{Bi_decomp}) into (\ref{b_rel})
we obtain the following linear system
\begin{eqnarray}
\mathcal{E}F^*_{i} & = & \sum^{r}_{j=1}X_j\mathcal{F}_{ji} + \sum^{r}_{j=1}\mathbf{C}X_j\mathcal{G}_{ji}
\label{fx_sys_multi}\\
\mathcal{F}_{ji} & = & \frac{\mu^*_i F^T_j\mathcal{E}F^*_i}{\mu_j(\mu_j - \mu^*_i)}\, ,\qquad
\mathcal{G}_{ji} = - \frac{\mu^*_i F^T_j\mathbf{C}\mathcal{E}F^*_i}{\mu_j(\mu_j + \mu^*_i)}
\end{eqnarray}
for factors $X_i$. Solving it allows one to express the factor $X_i$ through $F_i$.

\begin{ex}  This is especially easy to do when the dressing factor has a single pair of poles ($r=1$).
We shall drop subscripts for the sake of convenience. Then linear system (\ref{fx_sys_multi}) reduces to
\begin{equation}
\mathcal{E} F^{*} = \frac{\mu^*}{\mu}\left(X\frac{F^T\mathcal{E}F^*}{\mu - \mu^*}
-\mathbf{C}X \frac{F^T\mathbf{C}\mathcal{E}F^*}{\mu + \mu^*}\right) \cdot
\label{fx_sys}\end{equation}
Whenever $X$ and $F$ are column-vectors the result for $X$ reads:
\begin{equation}
X = \frac{\mu}{\mu^*}\left(\frac{F^T\mathcal{E}F^*}{\mu -\mu^*}
- \frac{F^T\mathbf{C}\mathcal{E} F^*}{\mu + \mu^*}\mathbf{C} \right)^{-1}\mathcal{E}F^*. \qquad\Box
\label{x_f}
\end{equation}
\end{ex}

The matrix factors $F_i(x,t)$ can be expressed in terms of fundamental solutions to the bare linear problem.
Indeed \cite{varna13}, a more detailed analysis of (\ref{g_pde1}) shows that the following relation holds true:
\begin{equation}
F^T_i(x,t) = F^T_{i,0}\hat{\Psi}_0(x,t,\mu_i)
\label{f_k_psi_0}
\end{equation}
where $F^T_{i,0}$ are constants of integration. Now we have all the information required to construct the dressed
solution at some initial moment of time $t=0$. To recover its time evolution we need to determine $F_{i,0}$ as
functions of $t$. The latter are governed by the linear equations
\begin{equation}
\rmi\partial_tF^T_{i,0} - F^T_{i,0}f(\mu_i) = 0
\label{f_i0}
\end{equation}
for $f(\lambda)$ being the dispersion law of NLEE. Thus in order to recover the time evolution of the dressed solution
we can apply the following simple correspondence:
\begin{equation}
F^T_{i, 0}\to F^T_{i,0}\,\rme^{-\rmi f(\mu_i)t}.
\label{f_k_prol_t}
\end{equation}

\subsection{Degenerate case I (real poles)}\label{dres2}

Let us now assume the poles of the dressing factor lie on the continuous spectrum of the operator $L$.
We shall consider first the case when all poles are real, i.e. $\mu^*_i = \mu_i$ for $i=1,\ldots,r$.
Then the identity $g(\lambda)\hat{g}(\lambda) = \openone$ gives rise to the algebraic relations:
\begin{eqnarray}
&& B_i\mathcal{E}B^{\dag}_i = 0\label{mum2_a3_r}\\
&& \Omega_i\mathcal{E}B^{\dag}_i\mathcal{E} + B_i\mathcal{E} \Omega^{\dag}_i\mathcal{E} = 0 \label{mum1_a3_r}
\end{eqnarray}
where
\begin{equation}
\Omega_i = \openone + \sum^{r}_{j\neq i}\frac{\mu_i B_j}{\mu_j(\mu_i - \mu_j)}
+ \sum^r_{j=1}\frac{\mu_i\mathbf{C}B_j\mathbf{C}}{\mu_j(\mu_i + \mu_j)}\,\cdot
\label{Omegai}
\end{equation}
It is straightforward from (\ref{mum2_a3_r}) that the residues are degenerate, i.e. the decomposition (\ref{Bi_decomp})
applies again. The matrices $F_i$ fulfill the quadratic relations:
\begin{equation}
F^T_i\mathcal{E}F^*_i = 0. 
\end{equation}
Equality (\ref{mum1_a3_r}) implies that there exist quadratic matrices $\alpha_i$ such that
\begin{equation}
\Omega_i\mathcal{E}F^*_i = X_i\alpha_i,\qquad \alpha^{\dag}_i = - \,\alpha_i.
\end{equation}
Thus we obtain the following linear system for $X_i$
\begin{eqnarray}
\mathcal{E}F^*_i &=& X_i\alpha_i - \mathbf{C}X_i\frac{F^T_i\mathbf{C}\mathcal{E}F^*_i}{2\mu_i}	\nonumber\\
 &+& \sum^r_{j\neq i} \left(X_j \frac{\mu_i F^T_j\mathcal{E}F^*_i}{\mu_j(\mu_j - \mu_i)}
 - \mathbf{C}X_j\frac{\mu_i F^T_j\mathbf{C}\mathcal{E}F^*_i}{\mu_j(\mu_j + \mu_i)}\right).
\label{alphafx_sys_multi}
\end{eqnarray}
By solving it we can express matrices $X_i$ in terms of $F_i$ and $\alpha_i$.

\begin{ex}
Like in the generic case discussed in the previous subsection this is especially easy when $g$ has a single pair
of poles and the rank of $X$ and $F$ equals 1. Then the linear system (\ref{alphafx_sys_multi}) simplifies to:
\begin{equation}
\mathcal{E}F^* = \left(\alpha -\frac{F^T\mathbf{C}\mathcal{E}F^*}{2\mu}\mathbf{C} \right)X
\label{alphafx_sys_sing}
\end{equation}
and the result for $X$ reads:
\begin{equation}
X = \left(\alpha -\frac{F^T\mathbf{C}\mathcal{E}F^*}{2\mu}\,\mathbf{C} \right)^{-1}\mathcal{E}F^* .\qquad\Box
\label{xf_a3_0}
\end{equation}
\end{ex}

In order to find $F_i$ and $\alpha_i$ we consider equation (\ref{g_pde1}). After evaluating the coefficients before powers of 
$\lambda - \mu_i$ we get the following differential relations:
\begin{eqnarray}
\rmi\partial_xF^T_i - F^T_iU^{(0)}(x,t,\mu_i) = 0 \label{diffm2_a3_x}\\ 
\rmi\partial_x\alpha_i - F^T_i\partial_{\lambda}|_{\lambda = \mu_i}U^{(0)} \mathcal{E}F^*_i = 0\label{diffm1_a3_x}
\end{eqnarray}
where $U^{(0)}(x,t,\lambda) = \lambda Q^{(0)}(x,t) - \lambda^2J$. Relation (\ref{diffm2_a3_x}) implies that $F_i$ is proportional
to a bare fundamental solution as given by (\ref{f_k_psi_0}). On the other hand, it can be shown that (\ref{diffm1_a3_x}) leads to
an interrelation between $\alpha_i$ and $F_i$, namely we have:
\begin{equation}
\alpha_i(x,t) = \alpha_{i,0}(t) - F^T_{i}(x,t)\partial_{\lambda}|_{\lambda = \mu_i}
\Psi_0(x,t,\lambda) \mathcal{E}F^*_{i,0}.
\label{alpha_a3_r}
\end{equation}
To find functions $F_{i,0}(t)$ and $\alpha_{i,0}(t)$ we have to consider equation (\ref{g_pde2}) this time.
As a result we derive the following relations:
\begin{eqnarray}
\rmi\frac{\rmd F^T_{i,0}}{\rmd t} - F^T_iV^{(0)}(x,t,\mu_i) = 0
\label{diffm2_a3_t}\\ 
\rmi\frac{\rmd \alpha_{i,0}}{\rmd t} - F^T_i\partial_{\lambda}|_{\lambda =
\mu_i}V^{(0)} \mathcal{E}F^*_i = 0.\label{diffm1_a3_t}
\end{eqnarray}
The former relation leads to exponential $t$-dependence of $F_{i,0}$ like in the generic case, see (\ref{f_i0}).
The second differential relation gives rise to:
\begin{equation}
\rmi\frac{\rmd \alpha_{i,0}}{\rmd t} - F^T_{i,0}\left. \frac{\rmd f(\lambda)}{\rmd\lambda}\right|_{\lambda = \mu_i}
\mathcal{E}F^*_{i,0} = 0
\label{alpha_i0}	
\end{equation}
which means that $\alpha_{i,0}$ is a linear function of time. Thus to recover the time dependence of the dressed
solution in this case one has to apply the rule given by (\ref{f_k_prol_t}) as well as:
\begin{equation}
\alpha_{i,0} \ \to\ \alpha_{i,0} - \rmi F^T_{i,0}\left. \frac{\rmd f(\lambda)}{\rmd\lambda}\right|_{\lambda = \mu_i}
\mathcal{E}F^*_{i,0} t .
\end{equation}

\subsection{Degenerate case II (imaginary poles)}\label{dres3}

Now let us suppose the poles of the dressing factor are all imaginary, i.e. $\mu^*_i = -\mu_i$, $i=1,\ldots,r$.
Then from the equality $g(\lambda)\hat{g}(\lambda) = \openone$ one obtains the following algebraic relations:
\begin{eqnarray}
&& B_i\mathcal{E}\mathbf{C}B^{\dag}_i = 0\label{mum2_a3_i}\\
&& B_i\mathcal{E}\mathbf{C}\Omega^{\dag}_i\mathbf{C}\mathcal{E} = \Omega_i\mathcal{E}\mathbf{C}B^{\dag}\mathbf{C}\mathcal{E}\label{mum1_a3_i}
\end{eqnarray}
where $\Omega_i$ is given by (\ref{Omegai}). As before the former algebraic relation means that $B_i(x,t)$ are degenerate matrices, hence
they are decomposed into a product of two rectangular matrices $X_i$ and $F_i$, see (\ref{Bi_decomp}). After substituting that decomposition
into (\ref{mum2_a3_i}) the latter gives rise to the quadratic relation:
\begin{equation}
F^T_i\mathcal{E}\mathbf{C}F^*_i = 0. 
\label{ff_a3_i}\end{equation}
Similarly to the previous case relation (\ref{mum1_a3_i}) is reduced to:
\begin{equation}
\Omega_i \mathcal{E}\mathbf{C}F^*_i = X_i\alpha_i 
\label{alpha_eq_a3_i}
\end{equation}
where we have that $\alpha^{\dag}_i = \alpha_i$. (\ref{alpha_eq_a3_i}) is viewed as a linear equation for $X_i$.
Solving it, allows one to express $X_i$ in terms of $F_i$ and $\alpha_i$.

\begin{ex}
Now consider the case when the dressing factor has a single pair of poles. Suppose $X(x,t)$ and $F(x,t)$ are $m+n$-vectors
and $\alpha$ is a real scalar function. Then (\ref{alpha_eq_a3_i}) is reduced to give
\begin{equation}
\mathcal{E}\mathbf{C}F^* = \left(\alpha - \frac{F^T\mathcal{E}F^*}{2\mu}\,\mathbf{C}\right) X.
\label{fx_a3}
\end{equation}
Clearly, the solution to (\ref{fx_a3}) is written down as follows:
\begin{equation}
X = \left(\alpha - \frac{F^T\mathcal{E}F^*}{2\mu} \,\mathbf{C}\right)^{-1}\mathcal{E}\mathbf{C}F^*.\qquad\Box
\label{xf_a3}
\end{equation}
\end{ex}

In order to find $F_i$ and $\alpha_i$ we again evaluate the coefficients before poles of the
equation (\ref{f_k_psi_0}). Thus we get the following equations:
\begin{eqnarray}
\rmi\partial_xF^T_i  & - & F^T_iU^{(0)} = 0 \qquad\Rightarrow\qquad\\
F^T_i(x,t) & = & F^T_{i,0}(t)\hat{\Psi}_0(x,t,\mu_i) \label{diffm2_a3}\\ 
\rmi\partial_x\alpha_i & - & F^T_i\partial_{\lambda}|_{\lambda = \mu_i}U^{(0)}\mathcal{E}\mathbf{C}F^*_i  = 0
\qquad\Rightarrow\qquad\nonumber\\
\alpha_i(x,t) & = & \alpha_{i,0}(t) - F^T_{i}(x,t)\dot{\Psi}_0(x,t,\mu_i)
\mathcal{E}\mathbf{C}F^*_{i,0}(t)
\label{alpha_a3_i}
\end{eqnarray}
where dot means differentiation in $\lambda$ and $U^{(0)}(x,t,\lambda)$ is the same as in the previous subsection. The time
dependence of $F_{i,0}$ and $\alpha_{i,0}$ is determined from equation (\ref{g_pde2}). The result reads:
\begin{eqnarray}
F^T_{i, 0} & \to & F^T_{i,0}\,\rme^{-\rmi f(\mu_i)t}\\
\alpha_{i,0} & \to & \alpha_{i,0} - \rmi F^T_{i,0}\left. \frac{\rmd f(\lambda)}{\rmd\lambda}\right|_{\lambda = \mu_i}
\mathcal{E}\mathbf{C}F^*_{i,0}\, t .
\end{eqnarray} 

\begin{rem}
We have discussed so far the situations when all poles of $g$ are either generic or belong to the continuous spectrum
of $L$. Apart from these "pure" cases one could consider a mixed one as well, i.e. part of poles are generic while the
rest are real or imaginary. Clearly, analysis of the mixed case is reduced to that of (some combination of) the
pure ones. $\Box$
\end{rem}

The algorithm to generate new solutions based on the dressing technique we described here can be symbolically
presented in the following diagram: 
\begin{eqnarray*}
Q_0\stackrel{(\ref{bare_x})}{\longrightarrow} \Psi_0\stackrel{(\ref{f_k_psi_0})}{\longrightarrow}\{F_j\}^{r}_{j=1}
\stackrel{(\ref{fx_sys_multi}), (\ref{alphafx_sys_multi}), (\ref{alpha_eq_a3_i})}{\longrightarrow}
\{X_j\}^{r}_{j=1}\stackrel{(\ref{Bi_decomp})}{\longrightarrow}\{B_j\}^{r}_{j=1} \stackrel{(\ref{q_1_q_0})}{\longrightarrow} Q_1.
\end{eqnarray*}

\section{Particular solutions}\label{sol}

We shall illustrate here the general considerations from the previous section by constructing reflectionless potentials
over zero background and extend them to solutions to DNLS (\ref{mdnls_pseudo}). Thus in what follows we shall assume that
$Q_0 = 0$. As a bare fundamental solution one can pick up the plane wave solution:
\begin{equation}
\Psi_0(x,t,\lambda) = \rme^{-\rmi\lambda^2 J x} .
\label{psi_seed}\end{equation}
Let us start with the case when the dressing factor has a single pair of complex poles in generic position.
Due to (\ref{f_k_psi_0}) and (\ref{psi_seed}) the rectangular factor $F$ is given by:
\begin{equation}
F(x,t) = \rme^{\rmi\mu^2 Jx}F(t),\qquad \mu\in\bbbc,\quad \mu^2\notin\bbbr.	
\label{f_psi_seed}
\end{equation}
Further on we shall restrict ourselves with the simplest case when $F(x,t)$ is simply a vector\footnote{$F_0$
is sometimes called polarization vector in theory of solitons.}. 
Taking into account (\ref{x_f}) and (\ref{f_psi_seed}) formula (\ref{q_1_q_0}) leads to the following
reflectionless potential:
\begin{eqnarray}
&\left(\mathbf{q}_{1}(x)\right)_{b a}& = \frac{2v}{\rho} 
\sum^{m+n}_{k=m+1}\frac{\epsilon_{a}\epsilon_{k}\epsilon_{b+m}\sin (2\varphi)
\rme^{-\rmi\sigma_{ak}(x)}\rme^{-\theta_{ak}(x)}}
{\Delta_k(x)}\nonumber\\
&\times &\left[\delta_{k b+m} -  \frac{2\rmi\epsilon_{b + m}\rme^{\rmi\gamma_{b + m \,k}}
\rme^{\xi_{b+m k}}\sin(2\varphi)}{\Delta_k(x)}\right]\\
&a = & 1, \ldots, m \qquad b = 1,\ldots, n \,. \nonumber
\label{refl_pot}
\end{eqnarray}
We have used above the polar representations 
\begin{equation}
\mu=\rho\exp(\rmi\varphi),\qquad F_{0,p} = |F_{0,p}|\exp(\rmi\phi_p)
\end{equation}
of the pole of $g$ and the components of polarization vector $F_0$ respectively as well as the following
auxiliary notations:
\begin{eqnarray*}
\Delta_k (x) & = & \rme^{-2\rmi\varphi}\sum^{m}_{p=1}\epsilon_{p}\rme^{-2\theta_{pk}(x)}
+ \sum^{m+n}_{p=m+1}\epsilon_{p}\rme^{2\xi_{pk}}\\
\theta_{pk}(x) &=& v x\sin(2\varphi) - \xi_{pk},\qquad \xi_{pk} = \ln|F_{0,p}/F_{0,k}|\\
\sigma_{pk}(x) &=& v x\cos(2\varphi) + \gamma_{pk} + \varphi,\qquad v = \frac{m + n}{m} \rho^2\\
\gamma_{pk} &=& \phi_{p} - \phi_{k} - 2\varphi.
\end{eqnarray*}
In order to obtain soliton type solution for the matrix DNLS (\ref{mdnls_pseudo}) one needs to recover the $t$-dependence in
(\ref{refl_pot}) using (\ref{f_k_prol_t}). The dispersion law for DNLS reads
\[f_{\mathrm{DNLS}}(\lambda) = -\frac{n+m}{m}\lambda^4J\, .\]
Thus one derives the following rule:
\begin{eqnarray}
\delta_{p} & \to & \delta_p + \left\{\begin{array}{ll}
v n\rho^2t\cos(4\varphi)/m ,&  p=1,\ldots, m\\
- v\rho^2t\cos(4\varphi) ,& p=m+1,\ldots, m+n \end{array}\right. \\
\xi_{pk} & \to &  \left\{\begin{array}{ll}
\xi_{pk} - v^2t\sin(4\varphi) ,& p=1,\ldots, m\\
\xi_{pk} ,& p=m+1,\ldots, m+n.\end{array}\right.
\end{eqnarray}
Let us consider a simple example.

\begin{ex}
The result we have just obtained represents a natural generalization of the soliton solution to the scalar DNLS
(\ref{dnls}) derived by Kaup and Newell in \cite{kaupnewel}. Indeed, for the simplest case when the Lax pair is related
to the Lie algebra $\mathfrak{sl}(2)$ we have $m = n = 1$ and $v = 2\rho^2$. Then $\mathbf{C} = \sigma_3$ and the dressing
factor (\ref{g_multi}) looks as follows:
\begin{equation}
g = \openone + \frac{\lambda B}{\mu(\lambda-\mu)} + \frac{\lambda\sigma_3B\sigma_3}
{\mu(\lambda+\mu)}\, .
\label{g_sl2}\end{equation}
According to (\ref{refl_pot}) the reflectionless potential can be written down as:
\begin{eqnarray}
q_1(x) &=&  \frac{4\rmi\rho\sin(2\varphi)\rme^{-\rmi\sigma(x)}
\rme^{\theta(x)}\left[\rme^{2\theta(x)}\pm \rme^{2\rmi\varphi} \right]}
{\left[\rme^{2\theta(x)}\pm \rme^{-2\rmi\varphi} \right]^2}\label{q_sl2}\\
\theta(x) &=& 2\rho^2x\sin(2\varphi) - \xi_{0}, \qquad \xi_{0} = \ln|F_{0,1}/F_{0,2}| \nonumber\\
\sigma(x) &=& 2\rho^2x\cos(2\varphi) - \phi_0, \qquad \phi_0 =\phi_2-\phi_1 - 3 \varphi.\nonumber
\end{eqnarray}
where the sign $\pm$ above refers to DNLS "$+$" or DNLS "$-$" respectively. To obtain the $1$-soliton
solution for (\ref{dnls}) we recover the time dependence in (\ref{q_sl2}) by using the correspondence:
\begin{equation}
\xi_0 \to \xi_0 - 4\rho^4t\sin(4\varphi),
\qquad\phi_0\to\phi_0 - 2\rho^4t\cos(4\varphi).
\label{t_recov_sl2}
\end{equation}
This way formulas (\ref{q_sl2})--(\ref{t_recov_sl2}) reproduce the soliton solution obtained by Kaup
and Newell by making use of Gelfand-Levitan-Marchenko equation. $\quad\Box$
\end{ex}

Let us consider now the degenerate case when $g$ has a single pair of real simple poles $\pm\rho$. Due to
(\ref{psi_seed}) formulas (\ref{f_k_psi_0}) and (\ref{alpha_a3_r}) give the following result for the vector
$F$ and the function $\alpha$
\begin{eqnarray}
F^T(x) &=& F^T_0 \rme^{\rmi\rho^2 Jx}\label{f_sol_a3_0}\\
\alpha(x) &=& \alpha_0 + \frac{2\rmi vx}{\rho}\sum^{m}_{p=1}\epsilon_{p}|F_{0,p}|^2.\label{alpha_sol_a3_0}
\end{eqnarray}
In order to obtain a solution to DNLS one needs to recover time evolution making use of the following
correspondence:
\begin{equation}
\begin{split}
F^T_0 &\to F^T_0\rme^{\rmi v\rho^2J t}\\
\alpha_0 &\to \alpha_0 + \frac{4\rmi v^2t}{\rho}\sum^{m}_{p=1}\epsilon_p|F_{0,p}|^2.
\end{split}
\end{equation}
Thus after taking into account (\ref{xf_a3_0}), (\ref{f_sol_a3_0}) and (\ref{alpha_sol_a3_0}) and set
$\alpha_0 =0$ formula (\ref{q_1_q_0}) gives rise to the following rational solution of the matrix DNLS
(\ref{mdnls_pseudo}):
\begin{eqnarray}
&\left(\mathbf{q}_{1}(x,t)\right)_{ba}& = \frac{2v}{\rho} \sum^{m+n}_{k=m+1}\frac{\epsilon_a\epsilon_k\epsilon_{b+m}
|\mathcal{F}_{a}\mathcal{F}_{k}|\rme^{-\rmi v\left(x + v t - \varphi_{a k}\right)}}
{2\rmi v \left(x + 2v t\right) -1}\nonumber\\
&\times& \left\lbrace \delta_{k\, b+m} - \frac{2\epsilon_{b + m} |\mathcal{F}_{b+m} \mathcal{F}_{k}|
\rme^{\rmi v \varphi_{k\,b+m}}}{2\rmi v \left(x + 2v t\right) -1}\right\rbrace \label{q_1f1_a3}
\\
&\varphi_{a b} = & (\arg \mathcal{F}_b - \arg\mathcal{F}_a)/v \nonumber
\end{eqnarray}
where we have used a normalized polarization vector:
\begin{equation}
\mathcal{F}_{s} = \frac{F_{0,s}}{\sqrt{\sum^m_{p=1}\epsilon_p|F_{0,p}|^2}},\qquad s=1,\ldots,m+n.
\label{f_norm}
\end{equation}

\begin{ex}
Suppose again we have a Lax pair associated with $\mathfrak{sl}(2)$, i.e. $m = n = 1$ and set $\mathcal{E} = \sigma_3$
(the choice $\mathcal{E} = \openone$ leads to a trivial result). Then the dressing factor is again given
by (\ref{g_sl2}) and the solution (\ref{q_1f1_a3}) simplifies to 
\begin{equation}
q_{1}(x,t) = 4\rho\frac{\left[1 + 4\rmi\rho^2\left(x + 4\rho^2t\right)\right]^3}
{\left[1 + 16\rho^4\left(x + 4\rho^2t\right)^2\right]^2}
\,\rme^{-2\rmi\rho^2 \left(x + 2\rho^2t\right)}. \qquad \Box
\label{q_1f1_sl2}
\end{equation}
\end{ex}
It is seen that (\ref{q_1f1_a3}) (and \ref{q_1f1_sl2}) is a not traveling wave solution and has no singularities.

Finally let us consider the case when the dressing factor has a pair of imaginary poles, i.e.
we assume $\mu = \rmi\rho$. Then the vector $F$ and the function $\alpha$ are given by: 
\begin{eqnarray}
F^T(x) &=& F^T_0 \rme^{-\rmi\rho^2 Jx}\label{f_sol_a3}\\
\alpha(x) &=& \alpha_0 - \frac{2 v x}{\rho} \sum^{m}_{p=1}\epsilon_{p}|F_{0,p}|^2.\label{alpha_sol_a3}
\end{eqnarray}
In order to obtain a solution to DNLS one needs to recover time evolution making use of the following
correspondence:
\begin{equation}
\begin{split}
F^T_0 & \to  F^T_0\rme^{\rmi v\rho^2J t}\\
\alpha_0 & \to \alpha_0 + \frac{4v^2t}{\rho}\sum^{m}_{p=1}\epsilon_p|F_{0,p}|^2.
\end{split}
\end{equation}
Thus after taking into account (\ref{xf_a3}), (\ref{f_sol_a3}) and (\ref{alpha_sol_a3}) and set
$\alpha_0 =0$ we get the following rational solution of the matrix DNLS:
\begin{eqnarray}
&\left(\mathbf{q}_{1}(x,t)\right)_{b a}& = \frac{2v}{\rho} \sum^{m+n}_{k=m+1}\frac{\epsilon_a\epsilon_k\epsilon_{b+m}
|\mathcal{F}_{a}\mathcal{F}_{k}|\rme^{\rmi v \left(x - v t + \varphi_{ak}\right)}}
{1 + 2\rmi v \left(x - 2v t\right)}\nonumber\\
&\times&\left\lbrace \delta_{k\, b + m} - \frac{2\epsilon_{b + m} |\mathcal{F}_{b + m} \mathcal{F}_{k}|
\rme^{\rmi v \varphi_{k\,b + m}}}{1 + 2\rmi v \left(x - 2v t\right)}\right\rbrace 
\label{q_1f2_a3}
\end{eqnarray}
where $\varphi_{a b} = (\arg \mathcal{F}_b - \arg\mathcal{F}_a)/v$ and by $\mathcal{F}$ is
denoted the normalized polarization vector defined in (\ref{f_norm}).
 
\begin{ex}
Suppose again we have a Lax pair associated with $\mathfrak{sl}(2)$. The only meaningful situation
now is when $\mathcal{E} = \openone$. Then expression (\ref{q_1f2_a3}) simplifies to 
\begin{equation}
q_{1}(x,t) = 4\rho\frac{\left[1- 4\rmi\rho^2\left(x - 4\rho^2t\right)\right]^3}
{\left[1 + 16\rho^4\left(x - 4\rho^2t\right)^2\right]^2} \,\rme^{2\rmi\rho^2
\left(x - 2\rho^2t\right)}. \qquad\Box
\label{q_1f2_sl2}
\end{equation}
\end{ex}
The rational solutions (\ref{q_1f1_sl2}) and (\ref{q_1f2_sl2}) we have just obtained coincides
with those in \cite{kaupnewel} derived from the soliton solution (\ref{q_sl2}) by taking a long-wave limit.

\begin{rem}
Function (\ref{q_1f2_a3}) (and \ref{q_1f2_sl2}) represents a nonsingular solution that is not a traveling wave. In fact, 
(\ref{q_1f2_sl2}) can be obtained from (\ref{q_1f1_sl2}) (similarly, (\ref{q_1f2_a3}) from (\ref{q_1f1_a3}))
after substituting $\mu = \rmi\rho$. However, one should keep in mind that those satisfy different NLEEs.\qquad $\Box$
\end{rem}

\begin{rem}
Clearly, one can recursively apply the dressing procedure we have demonstrated here thus building a whole infinite
sequence of exact solutions to DNLS. \qquad $\Box$
\end{rem}

\section{Conclusions}

We have adapted Zakharov-Shabat's dressing technique to quadratic bundles related to symmetric spaces of the
series $\mathbf{A.III}$. This allowed us to establish an algebraic procedure for construction of reflectionless
potentials which give rise to solutions to NLEEs of the DNLS hierarchy provided time dependence is appropriately
recovered. To do this it suffices to use dressing factors with simple poles symmetrically located to coordinate
frame, see (\ref{g_multi}). As an illustration, we have considered in more detail the case when the dressing
factor has just a single pair of poles. Using such a factor one can easily obtain explicit formulas for the solutions
of (\ref{mdnls_pseudo}). Our results naturally generalize those obtained by Kaup and Newell \cite{kaupnewel} for the
scalar DNLS which can be constructed by using a dressing factor in the form (\ref{g_sl2}). 

The dressing procedure developed in Section \ref{dres} naturally leads to two different classes of solutions:
generic soliton type of solutions (\ref{refl_pot}) and rational solutions, see (\ref{q_1f1_a3}) and
(\ref{q_1f2_a3}). In contrast to solutions of nonlinear Schr\"odinger equation like the following one
\[q(x,t) = \frac{\rme^{- 2\rmi\mu(x + 2\mu t)}}{x + 4\mu t},\qquad \mu\in\bbbr,\]
fast decaying rational solutions to DNLS are non-singular. The interest on rational solutions has significantly increased
\cite{deg1, deg2, hone} after it was observed that rogue waves in open ocean could be modeled through rational solutions to
nonlinear Schr\"odinger equation \cite{ahm1, ahm2, anc, per}. There is certain evidence\cite{deg1, rud2} that similar phenomena
in other media (like optical waveguides or plasma) could be described by rational solutions or solutions over nontrivial
background to other NLEEs as well.

Like in the scalar case, one could derive rational solutions (\ref{q_1f1_a3}) and (\ref{q_1f2_a3}) from (\ref{refl_pot}) through
a limiting procedure. Clearly, one could derive more complicated rational solutions through a similar limiting procedure but applied
on more complicated generic solutions, i.e. those constructed by using dressing factors with multiple pole pairs
or a recursive dressing by several single-pair factors. A major drawback of this approach, however, is that finding
generic solutions could lead to quite complicated calculations. This is where the procedures exposed in Subsection
\ref{dres2} and Subsection \ref{dres3} come into play. Those allow one to directly construct more complicated rational
type solutions without knowing the corresponding generic soliton type solutions.

Our results can be extended by constructing solutions over a non-trivial background. Such solutions were obtained in
\cite{dnlsratio, multisoldnls} for the case of the scalar DNLS. The considerations required in this
case are more complicated and we intend to discuss it elsewhere.

Another meaningful direction of further developments is to study quadratic bundles associated with other types
Hermitian symmetric spaces or, to put it even in a more general context, complete quadratic bundles related to
homogeneous spaces like the one given below:
\begin{equation}
L(\lambda) =  \rmi\partial_x + U_0 + \lambda U_1 - \lambda^2 J\, ,
\end{equation}
where $U_0$ splits into a diagonal and off-diagonal part, $U_1$ is strictly off-diagonal and $J$ is a diagonal
matrix. The theory of complete quadratic bundles like this one is more complicated than in the
case we have considered in that report. The latter represents certain interest in relation to N-wave type
equations with cubic non-linearity recently derived by Gerdjikov \cite{rhp}.

\section*{Acknowledgments}
The author would like to thank Dr Rossen Ivanov for useful discussions and support. The author would
also like to acknowledge financial support from the Government of Ireland Postdoctoral Fellowship in Science,
Engineering and Technology.

\end{document}